\documentclass[11pt, a4paper]{article}

\usepackage[utf8]{inputenc}
\usepackage[T1]{fontenc}
\usepackage{lmodern}                  
\usepackage[margin=1in, top=0.8in]{geometry}

\usepackage{amsmath, amssymb, amsthm}
\usepackage{graphicx}
\usepackage{hyperref}
\usepackage{natbib}   
\usepackage{esint}

\usepackage{authblk}                  
\usepackage{setspace}                 

\newcommand{\email}[1]{\texttt{#1}}

\title{Polarized image of an equatorial emitting ring around a Konoplya-Zhidenko rotating non-Kerr black hole}

\author[1,2]{Xin Qin\thanks{Corresponding author: \email{qx@hnust.edu.cn}}}
\author[3]{Fen Long\thanks{\email{lf@usc.edu.cn}}}
\author[4,5]{Songbai Chen\thanks{\email{csb3752@hunnu.edu.cn}}}
\author[4,5]{Jiliang Jing\thanks{\email{jljing@hunnu.edu.cn}}}
\affil[1]{\small{School of Physics and Electronic Science, Hunan University of Science and Technology, Xiangtan 411021, People’s Republic of China}}
\affil[2]{\small{Key Laboratory of Intelligent Sensors and Advanced Sensing Materials of Hunan Province, Hunan University of Science 
and Technology, Xiangtan 411021,People’s Republic of China}}
\affil[3]{\small{School of Mathematics and Physics, University of South China, Hengyang, 421001, People’s Republic of China}}
\affil[4]{\small{Department of Physics, Institute of Interdisciplinary Studies, Synergetic Innovation Center for Quantum Effects and Applications, Hunan Normal University, Changsha 410081, Hunan, People's Republic of China}}
\affil[5]{\small{Center for Gravitation and Cosmology, College of Physical Science and Technology, Yangzhou University, Yangzhou 225009, People's Republic of China}}

\date{}

\begin{document}

\maketitle

\begin{abstract}
We investigate the polarized images of an equatorial emitting ring around a Konoplya-Zhidenko rotating non-Kerr black hole, which introduces an additional deformation parameter. The deformation parameter $\eta$ allows the spin parameter to extend beyond the bounds imposed by the standard Kerr black hole. The results indicate that the polarized images depend not only on the magnetic field configuration, fluid velocity, and observer inclination angle, but also on the deformation parameter and the spin parameter. As the deformation parameter increases, the polarization intensity decreases monotonically. However, the magnitude of the Electric Vector Position Angle (EVPA) increases with $\eta$. Furthermore, we note that the parameter $\eta$ may induce subtle yet discernible azimuthal separation features, which could potentially distinguish it from the spin parameter and the magnetic field orientation angle. Nevertheless, these features are difficult to resolve under current observational conditions and await verification by future high-resolution facilities such as the next-generation Event Horizon Telescope (ngEHT).

\end{abstract}

\newpage
\clearpage

\section{Introduction}
The Event Horizon Telescope (EHT) Collaboration has released images of the supermassive black holes at the centers of M87 and Sgr A* \cite{EHT1,EHT2,EHT3,EHT4,EHT5,EHT6,EHT7}, marking the dawn of a new era in black hole astrophysics. Recently, polarized images of these black holes have also been published \cite{EHT8,EHT9,EHT10}, representing the first measurement in history of polarization information reflecting magnetic field properties in regions so close to their event horizons. These images exhibit strong and highly ordered spiral polarization patterns, suggesting that similar magnetic field configurations may exist in the vicinity of these supermassive black holes.
Studies indicate that the formation of polarization patterns depends not only on the physical properties of the accretion disk, but also on the curved spacetime structure near the black hole. Therefore, studies of black hole polarization images provide new insights into both the mechanisms underlying powerful jet formation and the physical properties of black holes. It also serves to test different theories of gravity.

Numerical simulations are typically employed to accurately render polarized images of black holes. However, such simulations incur extremely high computational costs, primarily due to the need for extensive parameter surveys and the complex coupling between astrophysical and relativistic effects. In recent years, Narayan et al. proposed a simplified radiative fluid ring model located on the equatorial plane of a black hole to study the polarized images of Schwarzschild black holes \cite{PZ1}. This model is capable of reproducing the polarization features observed in the M87* images released by the Event Horizon Telescope. One notable feature of this model is its ability to decouple and analyze the individual effects of various physical factors on polarized images, such as the magnetic field configuration around the black hole, the velocity profile of the accretion flow, and the observation inclination angle. The model was subsequently extended to more realistic scenarios involving rotating Kerr black holes \cite{PZ2}, and studies show that the differences in simulated images of the M87 black hole under low-spin and high-spin conditions primarily stem from differences in accretion dynamics. Recently, the polarized images around black holes have been investigated  \cite{PZ3,PZ4,PZ5,PZ6,PZ7,PZ8,PZ9,PZ10,PZ11,PZ12,PZ13}.

Although Einstein's general relativity has successfully passed numerous observational and experimental tests, current experimental results are still insufficient to completely rule out alternative theoretical models. Among these, Johannsen and Psaltis proposed a rotating non-Kerr spacetime metric that deviates from the Kerr metric \cite{JP1}, with the aim of testing the no-hair theorem of black holes. The spacetime proposed by Johannsen and Psaltis incorporates an additional deformation parameter, besides the mass and spin parameters, to quantify its deviation from the standard Kerr black hole spacetime. Recently, Konoplya and Zhidenko proposed another rotating non-Kerr black hole metric with static deformation \cite{KZ1}, which can be regarded as an axisymmetric vacuum solution of a yet-unknown alternative theory of gravity. Since the event horizon radius does not depend on the polar angle, the horizon surface in this spacetime retains a spherical structure, similar to the Kerr case. Research has shown that for certain values of the deformation parameter, the quasinormal modes of the Konoplya-Zhidenko rotating non-Kerr black hole are identical to those in the Kerr black hole case. This result provides strong support for the validity of such deformed black hole models. Furthermore, observational constraints from quasi-periodic oscillations \cite{KZ2} and the iron line \cite{KZ3} provide additional support for the hypothesis that this class of black holes can describe real astrophysical black holes. The physical properties and observational signatures of this class of rotating non-Kerr black hole have been extensively explored \cite{KZ4,KZ5,KZ6,KZ7,KZ8}. This paper aims to investigate the polarized images in the spacetime of the Konoplya-Zhidenko rotating non-Kerr black hole, analyzing the influence of the deformation parameter on the polarization intensity and the electric vector position angle of the black hole.

The structure of this paper is as follows: Section 2 briefly introduces the Konoplya-Zhidenko rotating non-Kerr black hole spacetime and presents the formula for the observed polarization vector. Section 3 displays the polarized images of the equatorial emitting ring and analyzes the effects of the deformation parameter on the polarization intensity and the electric vector position angle (EVPA).

\section{Observed polarization field in a Konoplya-Zhidenko rotating non-Kerr black hole}
The Konoplya-Zhidenko rotating non-Kerr black hole solution was introduced in \cite{KZ1}, describing a rotating black hole geometry that deviates from the Kerr solution via an additional deformation parameter. In Boyer-Lindquist coordinates, the metric takes the form
\begin{eqnarray}\label{Metric01}
ds^2=-\frac{\Delta\rho^2}{\Xi}dt^2+\frac{\Xi\sin^2\theta}{\rho^2}(d\phi-\omega{dt})^2+\frac{\rho^2}{\Delta}dr^2
+\rho^2{d\theta^2},
\end{eqnarray}
with
\begin{align}\label{Metric02}
\Delta&=r^2-2Mr+a^2-\frac{\eta}{r},\\ 
\rho^2&=r^2+a^2\cos^2\theta,  \\ 
\omega&=\frac{a(a^2+r^2-\Delta)}{\Xi}, \\
\Xi&=\left(r^2+a^2\right)^2-\Delta{a^2}\sin^2\theta.
\end{align}
where the parameters $M$, $a$ and $\eta$ correspond to the mass, spin, and deformation parameters, respectively. Compared to the Kerr black hole, the introduction of a deformation parameter not only extends the permissible range of the rotation parameter $a$ but also alters the spacetime structure in the strong-field region of the black hole spacetime. When $\eta=0$, the metric \eqref{Metric01} can reduce to the Kerr metric. In accordance with the weak cosmic censorship conjecture, we constrain our analysis to spacetime where the event horizon exists, ensuring that Equation (1) describes a black hole geometry. The condition for the existence of the event horizon is given by: 
\begin{equation}\label{yr_boudary}
\left\{
  \begin{array}{ll}
   \eta>0, \;\;\;\;\;\;\;\;(a>M)  \\
    \eta>{-\frac{2}{27}}\left(\sqrt{4M^2-3a^2}+2M\right)^2\left(\sqrt{4M^2-3a^2}-M\right), \;\;\;\;\;(a<M) \,
  \end{array}
\right.
\end{equation}
If the parameters $\eta$ and $a$ lie in other ranges, the spacetime describes a naked singularity.

To study the polarized images of the equatorial emitting ring in the Konoplya-Zhidenko rotating non-Kerr black hole spacetime, we first present the geodesic equation for photons in this spacetime:
\begin{eqnarray}\label{geodesic}
&&\frac{\rho^2}{E}p^t=\frac{r^2+a^2}{\Delta}\left(r^2+a^2-a\lambda\right)+a\left(\lambda-a\sin^2\theta\right), \\ \nonumber
&&\frac{\rho^2}{E}p^\phi=\frac{a}{\Delta}\left(r^2+a^2-a\lambda\right)+\frac{\lambda}{\sin^2\theta}-a, \\ \nonumber
&&\frac{\rho^2}{E}p^r=\pm_r\sqrt{\mathcal{R}(r)}, \\ \nonumber
&&\frac{\rho^2}{E}p^\theta=\pm_\theta\sqrt{\Theta(\theta)},
\end{eqnarray}  
where $\mathcal{R}(r)$ and $\Theta(\theta)$ representing the radial and angular potentials, respectively, which take the form
\begin{eqnarray}\label{potential} 
&&\mathcal{R}(r)=\left(r^2+a^2-a\lambda\right)^2-\Delta\left[\eta+(a-\lambda)^2\right],   \nonumber \\
&&\Theta(\theta)=\eta+a^2\cos^2\theta-\lambda^2\cot^2\theta.
\end{eqnarray}
The radial integral $I_r$ and the angle integral $G_\theta$ or a photon propagating from an initial position $(r_s,\theta_s=\frac{\pi}{2})$  along a null geodesic to a final position $(r_o\rightarrow\infty,\theta_o)$ are expressed as \cite{Gra1,Him1,WPC}
\begin{eqnarray}\label{integra01}
I_r\equiv\fint_{r_s}^{r_o}\frac{dr}{\pm_r\sqrt{\mathcal{R}(r)}}=\fint_{\theta_s}^{\theta_o}\frac{d\theta}{\pm_r\sqrt{\Theta(\theta)}}\equiv{G_\theta}.
\end{eqnarray}
where the slash through the integral sign indicates that the sign of $\pm_r$ or $\pm_\theta$ reverses when the photon passes through a radial or angular turning point. For a photon trajectory containing $m$ turning points, the radial integral is expressed as:
\begin{eqnarray}\label{xy1}
G_\theta^m=\frac{1}{\sqrt{-u_{-}a^2}}\left(2mK\left(\frac{u_+}{u_-}\right)-sign(y)F_o\right).
\end{eqnarray}
Moreover, the relationship between $(\lambda,\eta)$ and $(x,y)$ is expressed in the following form
\begin{eqnarray}\label{xy}
x=-\frac{\lambda}{\sin\theta_o}, \quad\quad\quad  y=\pm_o\sqrt{\Theta(\theta)}.
\end{eqnarray}
Based on Eqs.\eqref{integra01}, Eqs.\eqref{xy1} and Eqs.\eqref{xy}, a set of celestial coordinates $(x,y)$ at the observer's position can be obtained through numerical computation. 

Consider a point source on a radiation ring in the black hole's equatorial plane. In the local orthonormal frame of a zero-angular-momentum observer (ZAMO) at the source, the emitter's velocity vector lies entirely in the $\hat{r}-\hat{\phi}$ plane and takes the form:
\begin{eqnarray}\label{boost1}
\vec{\beta}=\beta_\nu\left[\cos\chi\left(\hat{r}\right)+\sin\chi(\hat{\phi})\right].
\end{eqnarray}
The photon's four-momentum in the boosted orthonormal frame can be obtained 
\begin{eqnarray}\label{trans02}
p^{(a)}=\Lambda^{(a)}_{\;\;\;(b)}\eta^{(b)(c)}e^\mu_{\;\;(c)}p_\mu.
\end{eqnarray}
where $e^\mu_{\;\;(c)}$ and $\Lambda^{(a)}_{\;\;\;(b)}$ are the zero-angular-momentum-observer tetrad and the Lorentz transformation\cite{PZ2}, respectively. Therefore, in the boosted orthonormal frame, the polarization vector is given by the cross product of the momentum $\vec{p}$ and the magnetic field $\vec{B}$, and can be expressed as:
\begin{eqnarray}\label{pv1}
&&f^{(t)}=0, \quad\quad\quad\quad\quad\quad f^{(r)}=\frac{p^{(\phi)}\times{B^{(\theta)}}}{|\vec{p}|}, \\  \nonumber
&&f^{(\theta)}=\frac{p^{(r)}\times{B^{(\phi)}}}{|\vec{p}|}, \quad\quad f^{(\phi)}=\frac{p^{(\theta)}\times{B^{(r)}}}{|\vec{p}|}.
\end{eqnarray}
In the fluid frame, the form of the magnetic field can be written as:
\begin{eqnarray}
&&\vec{B}=B_r\hat{e}_r+B_{\phi}\hat{e}_{\phi}+B_{\theta}\hat{e}_{\theta} \\ \nonumber
&&=B_{eq}(cos{\epsilon}\hat{e}_r+sin{\epsilon}\hat{e}_{\phi})+B_{\theta}\hat{e}_{\theta} \\ \nonumber
&&=\vec{B}_{eq}+B_{\theta}\hat{e}_{\theta}.
\end{eqnarray}
Thus, via the inverse transformation, the polarization four-vector $f^{\mu}$ can be expressed as
\begin{equation}\label{pv2}
f^\mu=e^\mu_{\;\;(b)}\Lambda_{(a)}^{\;\;(b)}f^{(a)},
\end{equation}
The polarization vector satisfies the normalization condition,
\begin{equation}
f^\mu{f_\mu}=\sin\zeta^2|\vec{B}|,
\end{equation}
where $\zeta$ is the angle between momentum $\vec{p}$ and magnetic field $\vec{B}$. Photons propagate along the geodesic, the polarization vector $f^{\mu}$ obeys
\begin{equation}
f^\mu{p_\mu}=0,\quad\quad\quad\quad p^\mu\nabla_\mu{f^\nu}=0.
\end{equation}

Since the Konoplya-Zhidenko rotating non-Kerr black hole (\eqref{Metric01}) belongs to a type D spacetime, the conserved Penrose-Walker constant $\kappa$ can be written as \cite{Chandrasekhar}
\begin{equation}\label{PW constant01}
\kappa=p^if_j(l_in_j-l_jn_i-m_i\bar{m}_j+\bar{m}_im_j)\Psi_2^{\left(-\frac{1}{3}\right)}.
\end{equation}
with
\begin{eqnarray}
&&\kappa=\kappa_1+i{\kappa_2}=(A-iB)\Psi_2^{\left(-\frac{1}{3}\right)},  \\ \nonumber
&&A=\left(p^tf^r-p^rf^t\right)+a\sin^2\theta\left(p^rf^\phi-p^\phi{f^r}\right),  \\ \nonumber
&&B=\left[\left(r^2+a^2\right)\left(p^\phi{f^\theta-p^\theta{f^\phi}}\right)-a\left(p^tf^\theta-p^\theta{f^t}\right)\right].
\end{eqnarray}
where $\Psi_2$ is the Weyl scalar and its explicit form is as follows
\begin{equation}
\Psi_2=\frac{a\cos\theta\left(5\imath{\eta}r-6\imath{M}r^3+a\eta\cos\theta\right)-2r^2\left(5\eta+3Mr^2\right)}{6r^3\left(r-\imath{a}\cos^3\theta\right)^3\left(r+\imath{a}\cos\theta\right)}.
\end{equation}

The Penrose-Walker conserved quantity serves as a bridge connecting the polarization states at the emission source and those measured by the observer. By combining the celestial coordinates $(x, y)$ with the Walker-Penrose conserved quantity at the source, the observed polarization vector can be expressed as
\begin{align}
f^x=\frac{y\kappa_2-\mu\kappa_1}{\mu^2+y^2},\quad\quad f^y=\frac{y\kappa_1+\mu\kappa_2}{\mu^2+y^2}, \quad\quad
\mu=-(x+a\sin\theta_o).
\end{align}
The observed intensity of linearly polarized synchrotron emission emitted by hot gas near a black hole is denoted by
\begin{eqnarray}\label{intens01}
|I|=g^{3+\alpha_\nu}l_p|\vec{B}|^{1+\alpha_\nu}(sin\zeta)^{1+\alpha_\nu},
\end{eqnarray}
where $g=\frac{E_o}{E_s}$ is the photon redshift between source and observer, and $l_p=\frac{p_s^{(t)}}{p_s^{(z)}}H$ is the geodesic path length through the emitting medium. The spectral index $\alpha_\nu=1$ is determined by the accretion disk properties. Thus, the observed polarization vector components are given by
\begin{eqnarray}\label{intens02}
f_{obs}^x=\sqrt{l_p}g^2|B|\sin\zeta f^x, \quad f_{obs}^y=\sqrt{l_p}g^2|B|\sin\zeta f^y.
\end{eqnarray}
The total polarized intensity and EVPA on the observer's screen are described by
\begin{eqnarray}\label{intens03}
I=\left(f_{obs}^x\right)^2+\left(f_{obs}^y\right)^2,  \quad\quad\quad \rm{EVPA}=\frac{1}{2}\arctan\frac{U}{Q},
\end{eqnarray}
Here, Q and U denote the Stokes parameters.
\begin{eqnarray}\label{intens04}
Q=\left(f_{obs}^y\right)^2-\left(f_{obs}^x\right)^2,  \quad\quad\quad U=-2f_{obs}^xf_{obs}^y.
\end{eqnarray}
In the Konoplya-Zhidenko rotating non-Kerr black hole spacetime, the polarization intensity and electric vector position angle of a point source can be computed using the celestial coordinates $(x, y)$ and Eqs.\eqref{PW constant01}, \eqref{intens01}–\eqref{intens04}. By repeating this procedure along the emitting ring, the influence of the deformation parameter on the overall polarization pattern can be demonstrated.
\begin{figure*}[htb!]
\centering
\includegraphics[width=15cm]{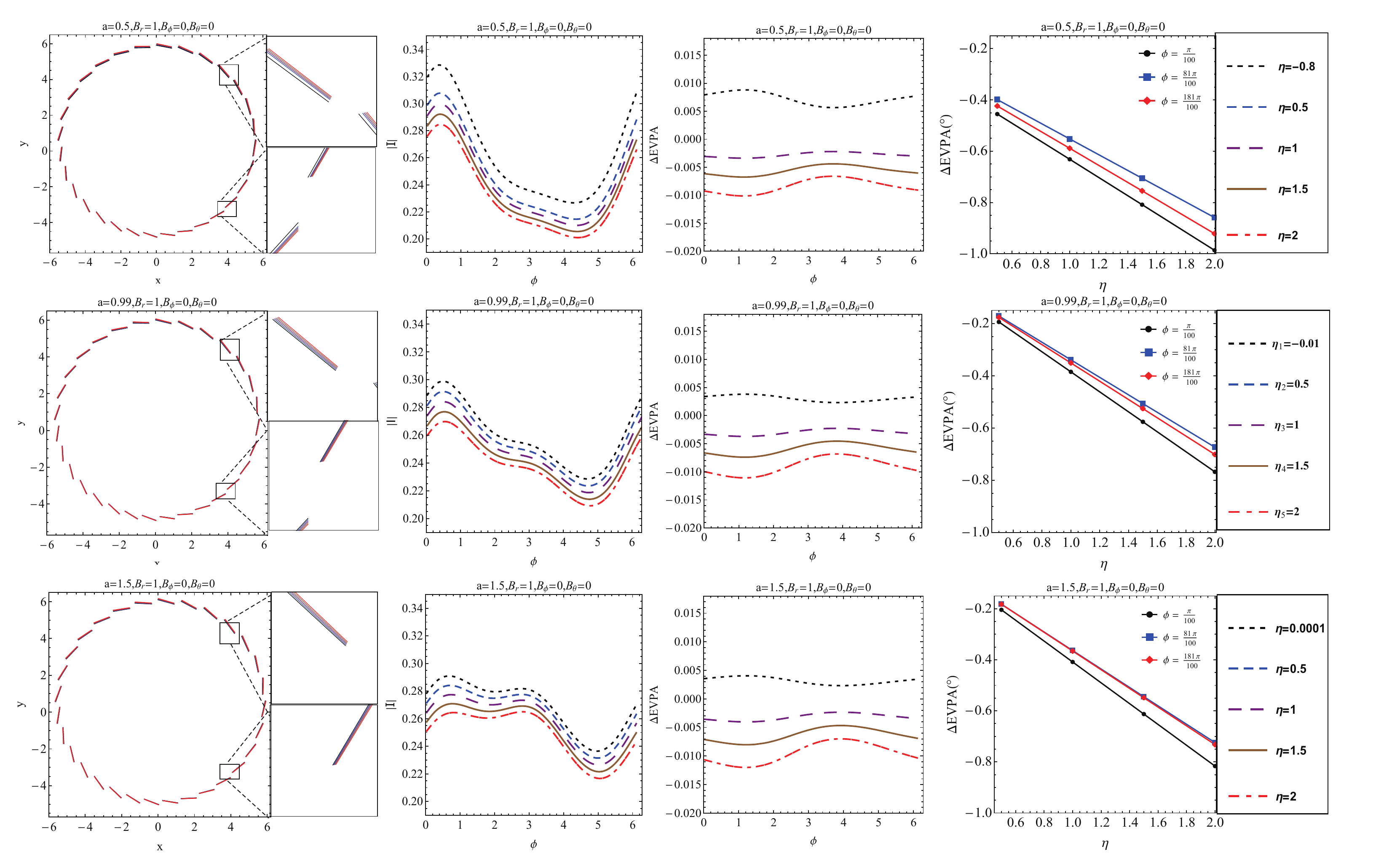}
\caption{Effects of $\eta$ on the polarization vectors and EVPA in the Konoplya-Zhidenko rotating non-Kerr black hole\eqref{Metric01} for different $a$, with the equatorial magnetic field with only an radial component $B_{r}$. Here $r_s=4.5$, $\theta_o=20^{\circ}$, $\beta_\nu=0.3$ and $\chi=-90^{\circ}$.}
\label{f1}
\end{figure*}
\begin{figure*}[htb!]
\centering
\includegraphics[width=15cm]{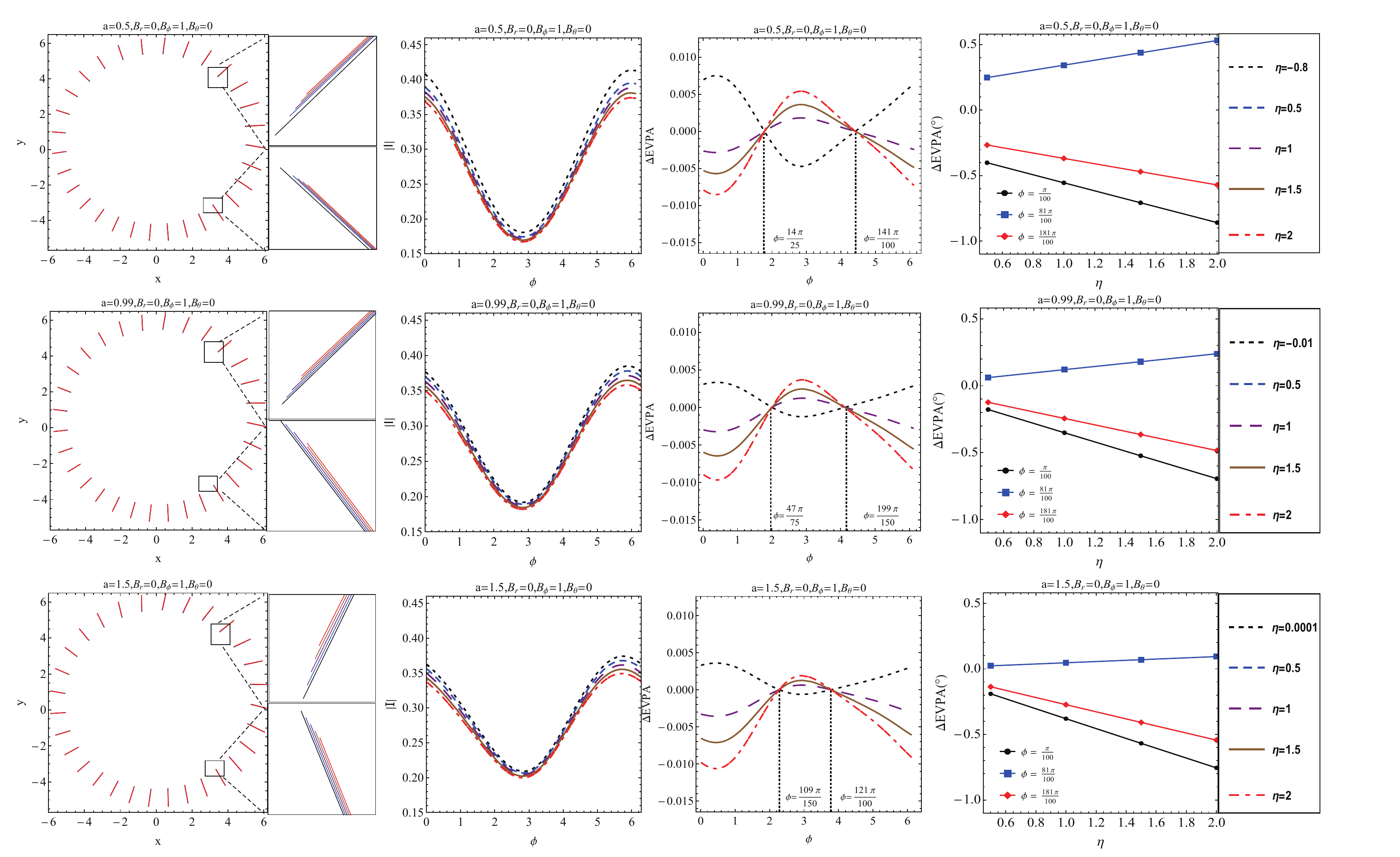}
\caption{Effects of $\eta$ on the polarization vectors and EVPA in the Konoplya-Zhidenko rotating non-Kerr black hole\eqref{Metric01} for different $a$, with the equatorial magnetic field with only an azimuthal component $B_{\phi}$. Here $r_s=4.5$, $\theta_o=20^{\circ}$, $\beta_\nu=0.3$ and $\chi=-90^{\circ}$.}
\label{f2}
\end{figure*}
\begin{figure*}[htb!]
\centering
\includegraphics[width=15cm]{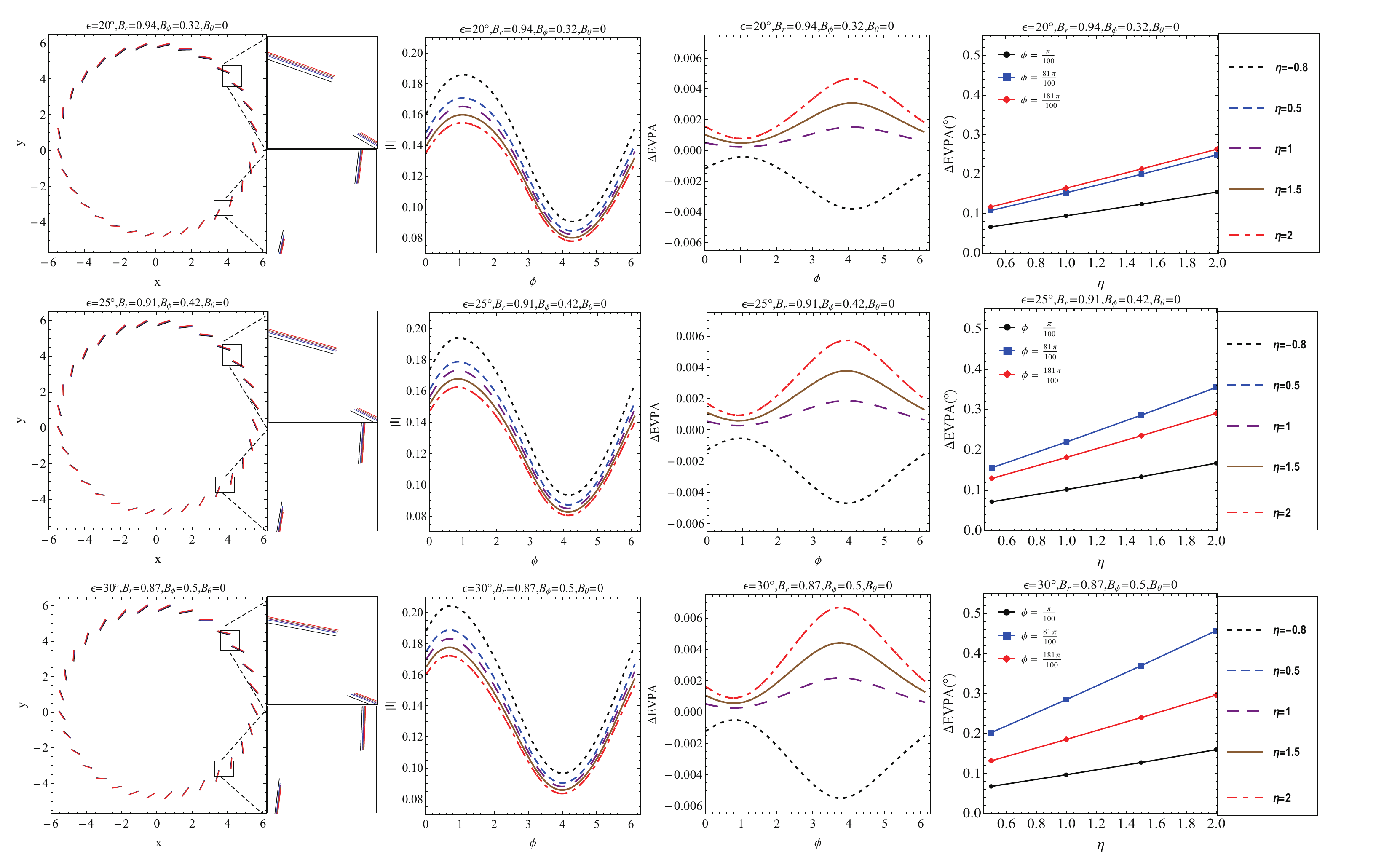}
\caption{Effects of $\eta$ on the polarization vectors and EVPA in the Konoplya-Zhidenko rotating non-Kerr black hole\eqref{Metric01} for different magnetic field. Here $r_s=4.5$, $a=0.5$ and $\beta_\nu=0.3$.}
\label{f3}
\end{figure*}

\section{Effects of the deformation parameter on the polarized image in the Konoplya-Zhidenko rotating non-Kerr black hole spacetime}

In this section, we present the polarized image of an emitting ring with a radius $r_s=4.5$ surrounding a Konoplya-Zhidenko rotating non-Kerr black hole. This ring radius is better matched to the observational characteristics of M87* \cite{PZ1,PZ2}. The results show that the pattern of the polarized image depends not only on the deformation parameters but is also influenced by factors such as the black hole's spin parameter, magnetic field structure, fluid velocity, and the observation inclination angle. The presence of the deformation parameter not only modifies the spacetime geometry around the black hole but also extends the allowed range of the spin parameter beyond that of the standard Kerr black hole. 
\begin{figure*}[htb!]
\centering
\includegraphics[width=15cm]{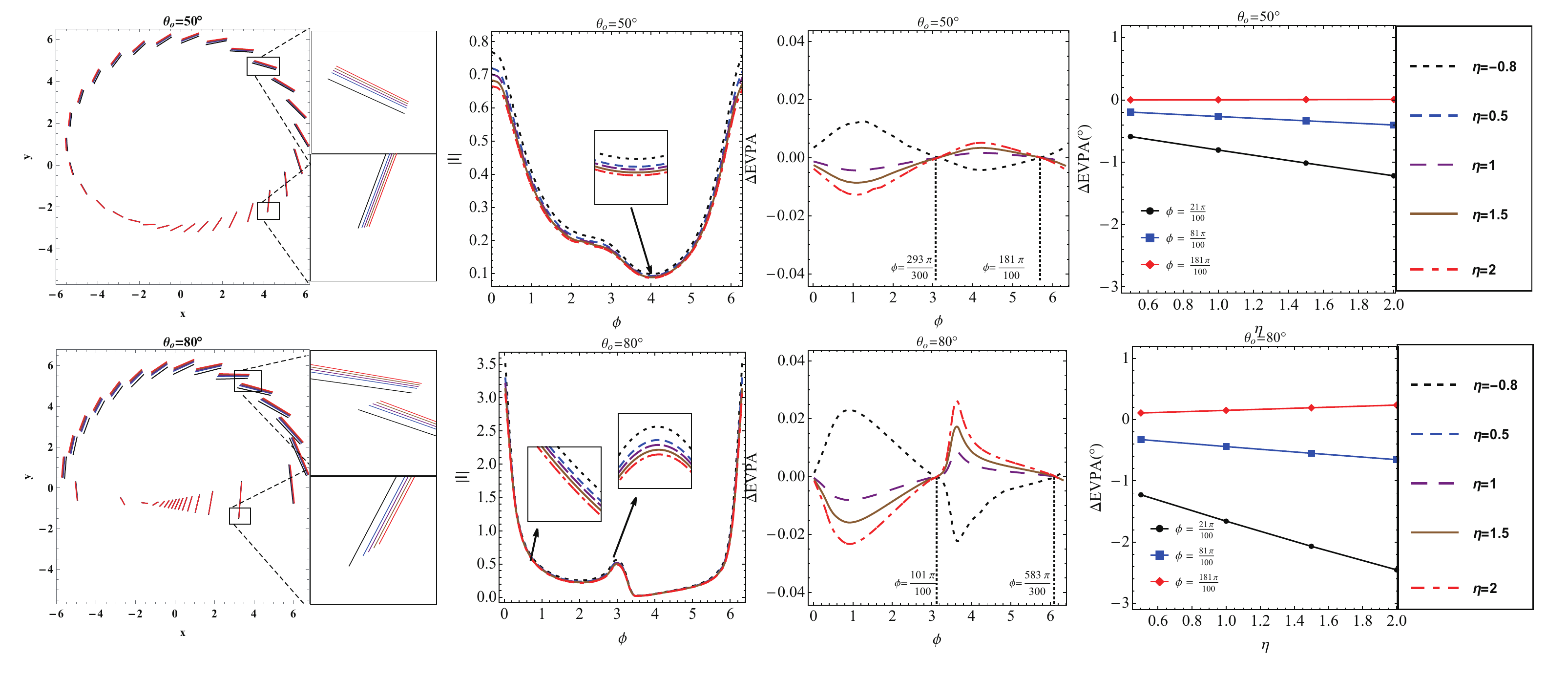}
\caption{Effects of $\eta$ on the polarization vectors and EVPA in the Konoplya-Zhidenko rotating non-Kerr black hole\eqref{Metric01} for different observer inclination angle $\theta_o$. Here $r_s=4.5$,  $a=0.5$, $\beta_\nu=0.3$, $\chi=-90^{\circ}$, $B_r=0.87$, $B_\phi=0.5$ and $B_\theta=0$.}
\label{f4}
\end{figure*}

\begin{figure*}[htb!]
\centering
\includegraphics[width=15cm]{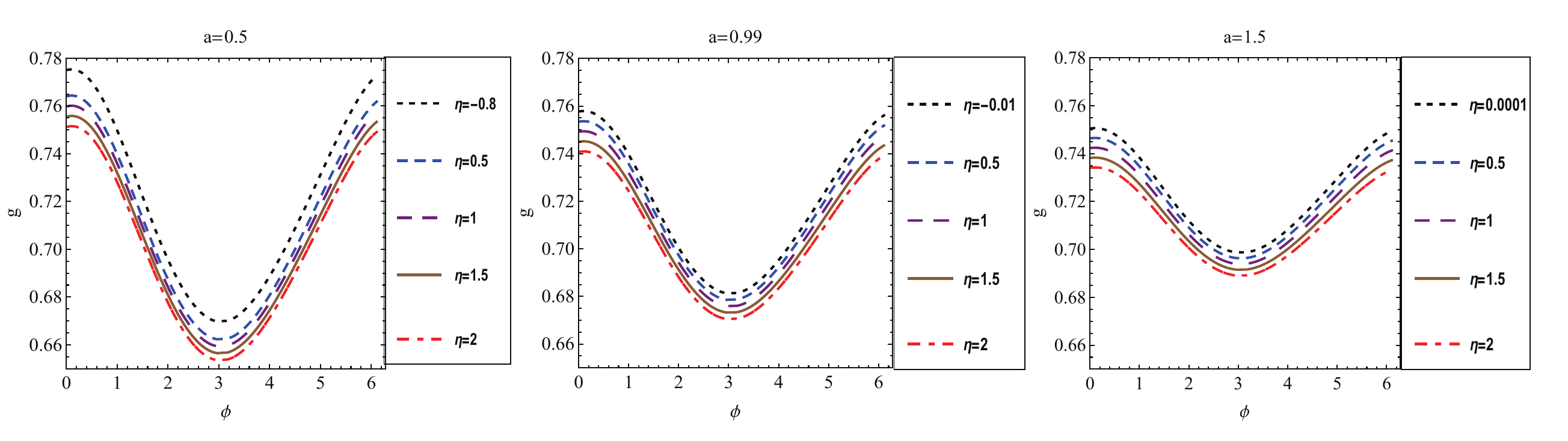}
\caption{Effects of $\eta$ on the redshift factor $g$ in the Konoplya-Zhidenko rotating non-Kerr black hole\eqref{Metric01}. Here $r_s=4.5$ and $\theta_o=20^{\circ}$.}
\label{f5}
\end{figure*}

We investigate the influence of the deformation parameter on the polarized images of the equatorial emission ring as the magnetic field lies in the equatorial plane in Figs.(\ref{f1})-(\ref{f4}). The quantity $\Delta{\rm EVPA}\equiv{\rm EVPA-EVPA_{\eta=0.5}}$. Figs.(\ref{f1})-(\ref{f4}) indicate that, for a fixed spin parameter $a$, the total polarized intensity $I$ decreases monotonically with increasing $\eta$. According to Eqs.\eqref{intens01} and \eqref{intens02}, the total polarized intensity is proportional to the fourth power of the redshift factor $g$. This strong dependence means that it is the dominant factor over other parameters. Consequently, Fig.(\ref{f5}) further investigates the impact of the deformation parameter on the redshift factor. The results demonstrate that the redshift factor exhibits a significant overall decrease as $\eta$ increases. This behavior can be attributed to the modification of the spacetime geometry by $\eta$. Specifically, $\eta$ enters the time component of the metric ($g_{tt}$) as a positive term, effectively reducing the absolute value of $g_{tt}$. This modification alters the time dilation effect, which in turn intensifies the gravitational redshift experienced by photons. Consequently, this leads to a decrease in the redshift factor $g$ and, subsequently, the total polarized intensity $I$.

However, the variation of the EVPA with respect to the deformation parameter also depends on the magnetic field configuration. As illustrated in Figs.(\ref{f1}) and (\ref{f2}), we observe that for a pure radial magnetic field, the EVPA decreases monotonically, whereas for a pure azimuthal magnetic field, it exhibits non-monotonic behavior. Although the trends differ across magnetic field structures, the fourth column reveals that the absolute value of the $\Delta$EVPA consistently increases with the deformation parameter.  This behavior arises because the polarization vector $\vec{f}^{\mu}$ undergoes parallel transport along null geodesics as photons propagate through curved spacetime. This implies that a greater degree of spacetime deformation results in a larger rotation of the polarization direction during propagation. 

Fig.(\ref{f3}) presents the polarized images under different magnetic field structures. As indicated in  \cite{PZ1}, the simulated images in the radially dominated magnetic field regime are more consistent with the actual observations of M87*. Therefore, we focus our study on the regime where $B_r>B_{\phi}$. We find that $\Delta$EVPA exhibits a monotonic increasing trend with the increase of $\eta$. Furthermore, for a fixed $\eta$ , the magnitude of $\Delta$EVPA generally increases with the magnetic field orientation angle. In Fig.(\ref{f4}), we further analyze the polarized images at different observer inclination angles. In the high-inclination regime, the polarization intensity decreases with increasing $\eta$. Additionally, the variation amplitude of the polarization position angle increases significantly with the increase of the observer inclination.

\begin{figure*}[htb!]
\centering
\includegraphics[width=13cm]{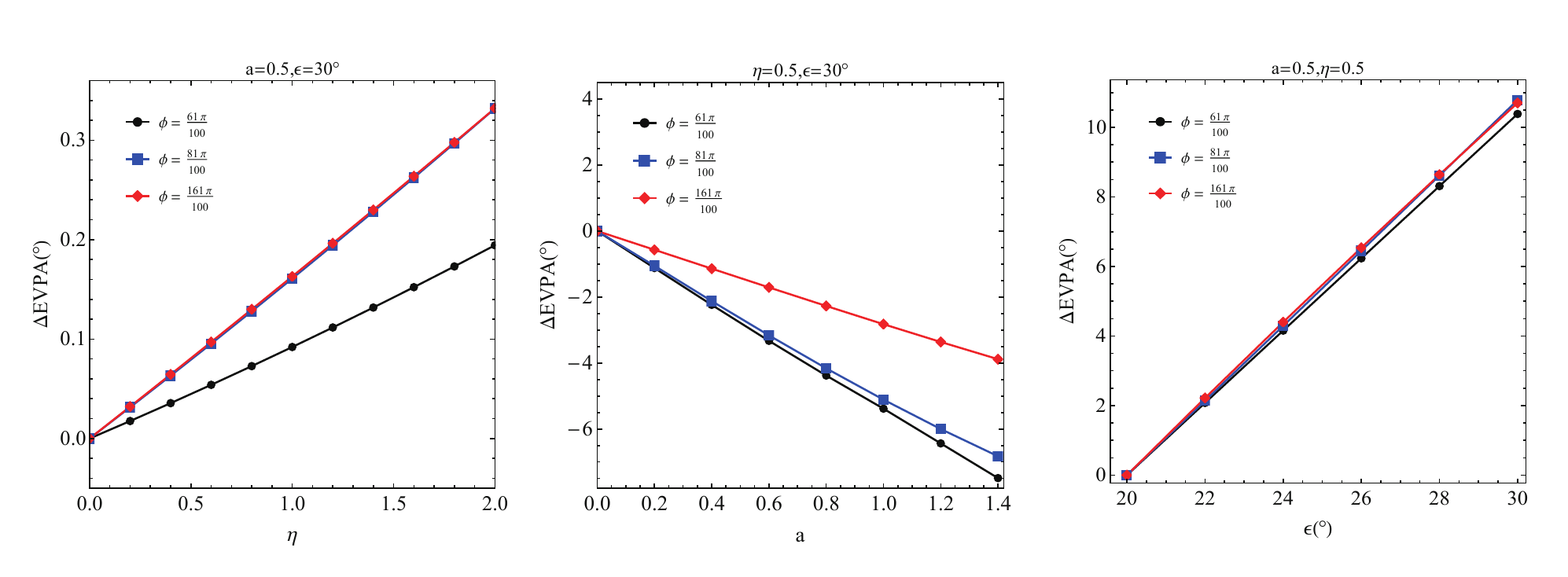}
\caption{Effects of the deformation parameter $\eta$ (left), the spin parameter $a$ (middle) and the magnetic field direction angle $\epsilon$(right) on the $\Delta$EVPA. Here $r_s=4.5$ and $\theta_o=20^{\circ}$.}
\label{f6}
\end{figure*}

Next, we investigate the dependence of $\Delta$EVPA on the deformation parameter $\eta$, spin parameter $a$ and magnetic field orientation angle $\epsilon$ at specific azimuthal angles, as illustrated in Fig.(\ref{f6}). The deformation parameter $\eta$ exhibits a positive correlation with $\Delta$EVPA. Notably, the curves for $\phi=81\pi/100$ and $\phi=161\pi/100$ nearly coincide, whereas the profile for $\phi=61\pi/100$ exhibits a distinct separation from the others, indicating a specific sensitivity to the azimuthal angle. In contrast, the spin parameter $a$ induces an overall negative trend in $\Delta$EVPA. A significant separation is observed in the red line ($\phi=161\pi/100$), suggesting that $a$ also possesses angular sensitivity, albeit distinct from that of $\eta$. Finally, the magnetic field direction angle $\epsilon$ leads to a rise in $\Delta$EVPA. The three lines representing different azimuthal angles remain closely clustered, a behavior distinctly different from the impact of $\eta$. Although the deformation parameter induces weakly distinguishable azimuthal separation features in $\Delta$EVPA, these subtle signatures are challenging to resolve with current observational capabilities. Future high-resolution facilities, such as the next-generation Event Horizon Telescope (ngEHT), are expected to detect these parameter-specific angular signals.
\begin{figure*}[htb!]
\centering
\includegraphics[width=14cm]{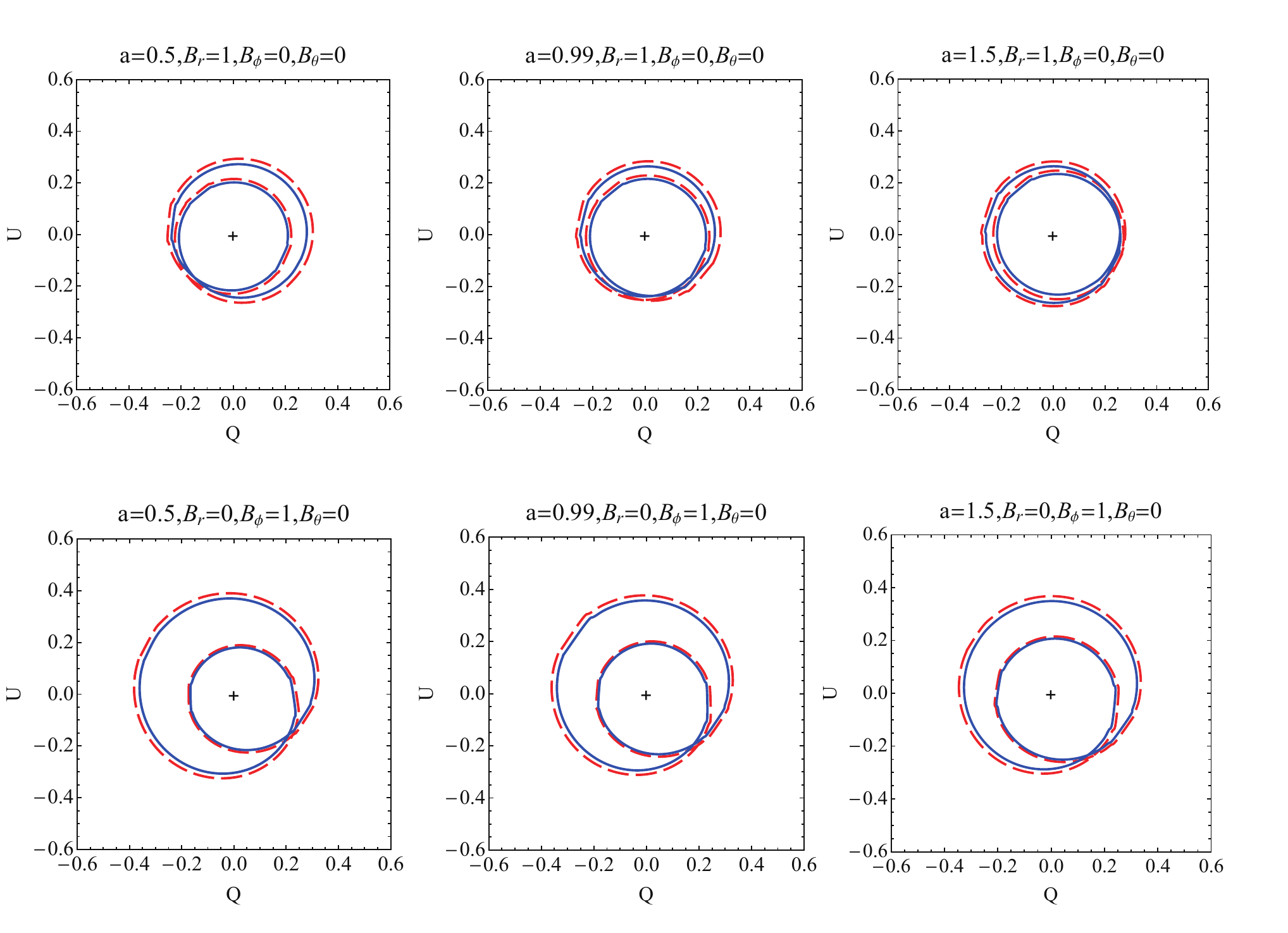}
\caption{Effects of $\eta$ on the $Q-U$ diagram in the Konoplya-Zhidenko rotating non-Kerr black hole \eqref{Metric01} for different $a$. Here $r_s=6$, $\theta_o=20^{\circ}$ and $\beta_\nu=0.3$. The red dashed line and the blue solid line represent the cases with $\eta=0.5$ and $\eta=2$, respectively. The black crosshairs denote the origin point for each plot.}
\label{f7}
\end{figure*}

\begin{figure*}[htb!]
\centering
\includegraphics[width=14cm]{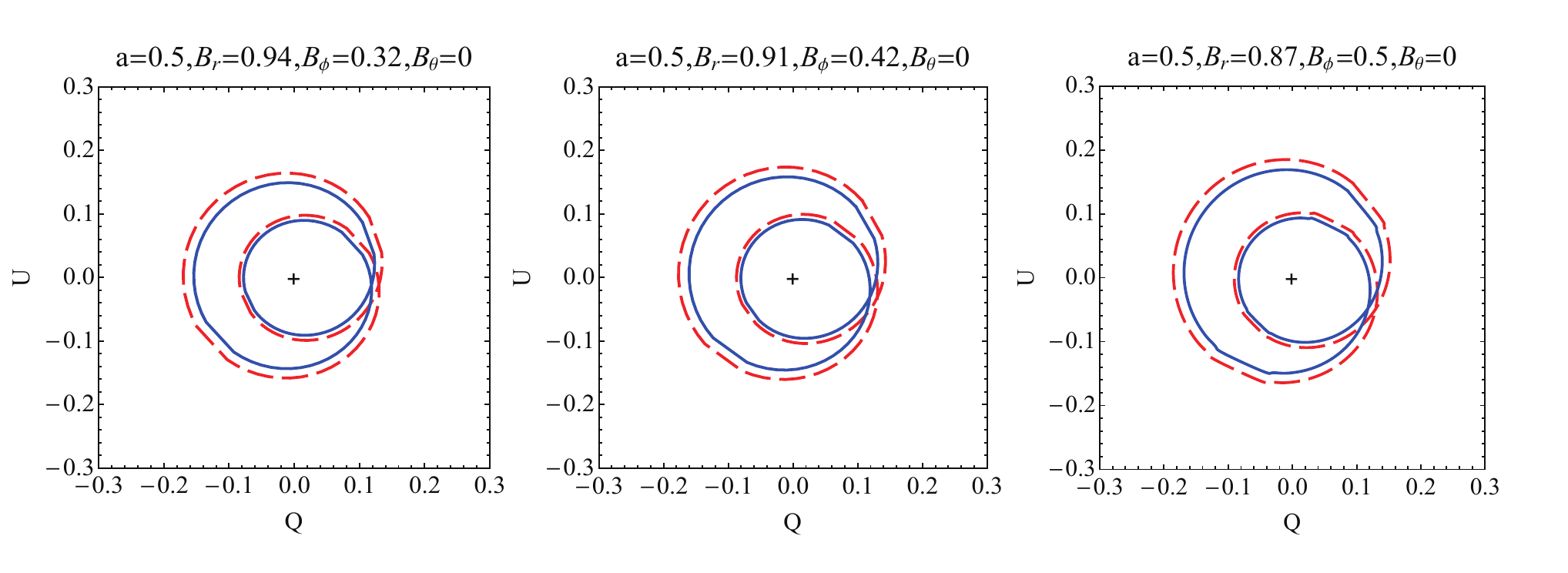}
\caption{Effects of $\eta$ on the $Q-U$ diagram in the Konoplya-Zhidenko rotating non-Kerr black hole \eqref{Metric01} for different magnetic field structures. Here $r_s=6$, $\theta_o=20^{\circ}$ and $\beta_\nu=0.3$. The red dashed line and the blue solid line represent the cases with $\eta=0.5$ and $\eta=2$, respectively. The black crosshairs denote the origin point for each plot.}
\label{f8}
\end{figure*}
Finally, we present the Stokes Q–U loops under different spin parameters and magnetic field configurations in Figs.(\ref{f7})-(\ref{f8}). The results indicate that, for a fixed observer inclination $\theta_o=20^{\circ}$ and specific magnetic field geometry, the sizes of both the outer and inner loops exhibit a contraction trend as the deformation parameter $\eta$ increases. Under the premise of a relatively simple magnetic field structure, a larger observed polarization loop tends to favor the hypothesis that the spacetime geometry approaches that of a standard Kerr black hole.

\section{Summary}

In this study, we investigate the polarized images of an equatorial emitting ring surrounding a Konoplya-Zhidenko rotating non-Kerr black hole. The results indicate that increasing the deformation parameter $\eta$ leads to a decrease in the total polarized intensity. Although the evolution of the EVPA depends on the magnetic field structure, the magnitude of its variation (|$\Delta$EVPA|) consistently increases with $\eta$. We further observe that for a fixed $\eta$, the magnitude of $\Delta$EVPA generally increases with the magnetic field orientation angle. Moreover, in the high-inclination regime, the polarization intensity decreases with increasing $\eta$, while the variation amplitude of the polarization position angle is found to increase significantly with the observer inclination.

We also analyze the effects of the deformation parameter $\eta$, the spin parameter $a$ and the magnetic field orientation angle $\epsilon$ on $\Delta$EVPA. The parameter $\eta$ exhibits a significant positive correlation with $\Delta$EVPA, and its influence exhibits a  unique azimuthal dependence. In contrast, the spin parameter $a$ leads to a general decrease in $\Delta$EVPA, with a markedly different azimuthal dependence from that of $\eta$. While $\epsilon$ increases the magnitude of $\Delta$EVPA, the curves corresponding to different azimuthal angles remain closely clustered. Consequently, the weak, yet discernible azimuthal separation induced by $\eta$ constitutes a unique signature that distinguishes it from the effects of $a$ and $\epsilon$, holding promise for detection by facilities such as the next-generation Event Horizon Telescope (ngEHT). Finally, we investigate the influence of $\eta$ on the Stokes Q-U loops. Our analysis reveals that both the outer and inner loop sizes shrink as $\eta$ increases. This implies that, assuming a relatively simple magnetic field configuration, a larger observed polarization loop favors a spacetime geometry closer to that of a standard Kerr black hole.

\section*{Acknowledgments}
This work was supported by the National Natural Science Foundation of China (Grant No.12405053, 12275078, 12205140, 11875026, 12035005, 2020YFC2201400), and the Natural Science Foundation of Hunan Province (Grant No. 2024JJ6211, 2023JJ40523), and the Scientific Research Fund of Hunan Provincial Education Department(Grant No. 24C0229).

\bibliography{references}     

@article{EHT1,
  author = {{Event Horizon Telescope Collaboration}},
  title = {First M87 Event Horizon Telescope Results. I. The Shadow of the Supermassive Black Hole},
  journal = {Astrophys. J. Lett.},
  year = {2019},
  volume = {875},
  pages = {L1}
}

@article{EHT2,
  author = {{Event Horizon Telescope Collaboration}},
  title = {First M87 Event Horizon Telescope Results. II. Array and Instrumentation},
  journal = {Astrophys. J. Lett.},
  year = {2019},
  volume = {875},
  pages = {L2}
}

@article{EHT3,
  author = {{Event Horizon Telescope Collaboration}},
  title = {First M87 Event Horizon Telescope Results. III. Data Processing and Calibration},
  journal = {Astrophys. J. Lett.},
  year = {2019},
  volume = {875},
  pages = {L3}
}

@article{EHT4,
  author = {{Event Horizon Telescope Collaboration}},
  title = {First M87 Event Horizon Telescope Results. IV. Imaging the Central Supermassive Black Hole},
  journal = {Astrophys. J. Lett.},
  year = {2019},
  volume = {875},
  pages = {L4}
}

@article{EHT5,
  author = {{Event Horizon Telescope Collaboration}},
  title = {First M87 Event Horizon Telescope Results. V. Physical origin of the asymmetric ring},
  journal = {Astrophys. J. Lett.},
  year = {2019},
  volume = {875},
  pages = {L5}
}

@article{EHT6,
  author = {{Event Horizon Telescope Collaboration}},
  title = {First M87 Event Horizon Telescope Results. VI. The Shadow and Mass of the Central Black Hole},
  journal = {Astrophys. J. Lett.},
  year = {2019},
  volume = {875},
  pages = {L6}
}

@article{EHT7,
  author = {{Event Horizon Telescope Collaboration}},
  title = {First Sagittarius A* Event Horizon Telescope Results. I. The Shadow of the Supermassive Black Hole in the Center of the Milky Way},
  journal = {Astrophys. J. Lett.},
  year = {2022},
  volume = {930},
  pages = {L12}
}

@article{EHT8,
  author = {{Event Horizon Telescope Collaboration}},
  title = {First M87 Event Horizon Telescope Results. VII. Polarization of the Ring},
  journal = {Astrophys. J. Lett.},
  year = {2019},
  volume = {875},
  pages = {L7}
}

@article{EHT9,
  author = {{Event Horizon Telescope Collaboration}},
  title = {First M87 Event Horizon Telescope Results. VIII. Magnetic Field Structure near the Event Horizon},
  journal = {Astrophys. J. Lett.},
  year = {2019},
  volume = {875},
  pages = {L8}
}

@article{EHT10,
  author = {{Event Horizon Telescope Collaboration}},
  title = {First M87 Event Horizon Telescope Results. IX. Testing the Kerr Metric},
  journal = {Astrophys. J. Lett.},
  year = {2019},
  volume = {875},
  pages = {L9}
}

@article{PZ1,
  author = {R. Narayan and D. C. M. Palumbo and M. D. Johnson and Z. Gelles and E. Himwich and D. O. Chang and A. Ricarte and J. Dexter and C. F. Gammie and A. A. Chael and {The Event Horizon Telescope Collaboration}},
  title = {The Polarized Image of a Synchrotron-emitting Ring of Gas Orbiting a Black Hole},
  journal = {Astrophys. J.},
  year = {2021},
  volume = {912},
  number = {1},
  pages = {35}
}

@article{PZ2,
  author = {Z. Gelles and E. Himwich and D. C. M. Palumbo and M. D. Johnson},
  title = {Polarized Image of Equatorial Emission in the Kerr Geometry},
  journal = {Physical Review D},
  year = {2021},
  volume = {104},
  number = {4},
  pages = {044060}
}

@article{PZ3,
  author = {X. Qin and S. Chen and J. Jing},
  title = {Polarized image of an equatorial emitting ring around a 4D Gauss-Bonnet black hole},
  journal = {Eur. Phys. J. C},
  year = {2022},
  volume = {82},
  pages = {784}
}

@article{PZ4,
  author = {Z. Hu and Y. Hou and H. Yan and M. Guo and B. Chen},
  title = {Polarized images of synchrotron radiations in curved spacetime},
  journal = {Eur. Phys. J. C},
  year = {2022},
  volume = {82},
  pages = {1166}
}

@article{PZ5,
  author = {X. Qin and S. Chen and Z. Zhang and J. Jing},
  title = {Polarized image of a rotating black hole surrounded by a cold dark matter halo},
  journal = {Eur. Phys. J. C},
  year = {2023},
  volume = {83},
  pages = {159}
}

@article{PZ6,
  author = {H. Zhu and M. Guo},
  title = {Polarized image of synchrotron radiations of hotspots in Schwarzschild-Melvin black hole spacetime},
  journal = {arxiv},
  year = {2023},
  pages = {2205.04777}
}

@article{PZ7,
  author = {Y. Hou and J. Huang and M. Guo and Y. Mizuno and B. Chen},
  title = {Near-horizon Polarization as a Diagnostic of Black Hole Spacetime},
  journal = {Astrophys. J. Lett.},
  year = {2025},
  volume = {988},
  pages = {L51}
}

@article{PZ8,
  author = {F. Zhou and J. Huang and Y. Li and Z. Zhang and Y. Hou and M. Guo and B. Chen},
  title = {Non-thermal synchrotron emission and polarization signatures during black hole flux eruptions},
  journal = {arxiv},
  year = {2025},
  pages = {2512.06803}
}

@article{PZ9,
  author = {V. Deliyski and G. Gulychev and P. Nedkova and S. Yazadjiev},
  title = {Polarized image of equatorial emission in horizonless spacetimes: Naked singularities},
  journal = {Phys. Rev. D},
  year = {2023},
  volume = {108},
  pages = {104049}
}

@article{PZ10,
  author = {S. Gu and Y. Huang and K. Liu and E. Liang and K. Lin},
  title = {Influence of quantum correction on the Schwarzschild black hole polarized image},
  journal = {Eur. Phys. J. C},
  year = {2024},
  volume = {84},
  pages = {601}
}

@article{PZ11,
  author = {H. Shi and T. Zhu},
  title = {Polarized image of a synchrotron emitting ring around a static hairy black hole in Horndeski theory},
  journal = {Eur. Phys. J. C},
  year = {2024},
  volume = {84},
  pages = {814}
}

@article{PZ12,
  author = {Y. Chen and L. Cheng and P. Wang and H. Yang},
  title = {Polarized Image of a synchrotron-emitting Ring in Einstein-Maxwell-scalar Theory},
  journal = {arxiv},
  year = {2024},
  pages = {2409.05304}
}

@article{PZ13,
  author = {C. Chen and Q. Pan and J. Jing},
  title = {Near-horizon polarized images of a rotating hairy Horndeski black hole},
  journal = {J. Cosmol. Astropart. P.},
 volume = {04},
  pages = {024}
}

@article{JP1,
  author = {T. Johannsen and D. Psaltis},
  title = {A Metric for Rapidly Spinning Black Holes Suitable for Strong-Field Tests of the No-Hair Theorem},
  journal = {Phys. Rev. D},
  year = {2011},
  volume = {83},
  pages = {124015}
}

@article{KZ1,
  author = {R. Konoplya and A. Zhidenko},
  title = {Detection of gravitational waves from black holes: Is there a window for alternative theories?},
  journal = {Phys. Lett. B},
  year = {2016},
  volume = {756},
  pages = {350}
}

@article{KZ2,
  author = {C. Bambi and S. Nampalliwar},
  title = {Quasi-periodic oscillations as a tool for testing the Kerr metric: A comparison with gravitational waves and iron line},
  journal = {Europhys. Lett.},
  year = {2020},
  volume = {130},
  pages = {10006}
}

@article{KZ3,
  author = {Y. Ni and J. Jiang and C. Bambi},
  title = {Testing the Kerr metric with the iron line and the KQ2 parametrization},
  journal = {J. Cosmol. Astropart. P.},
  year = {2016},
  volume = {09},
  pages = {014}
}

@article{KZ4,
  author = {S. Wang and S. Chen and J. Jing},
  title = {Strong gravitational lensing by a Konoplya-Zhidenko rotating non-Kerr compact object},
  journal = {J. Cosmol. Astropart. P.},
  year = {2016},
  volume = {11},
  pages = {020}
}

@article{KZ5,
  author = {M. Wang and S. Chen and J. Jing},
  title = {Shadow casted by a Konoplya-Zhidenko rotating non-Kerr black hole},
  journal = {J. Cosmol. Astropart. P.},
  year = {2017},
  volume = {01},
  pages = {051}
}

@article{KZ6,
  author = {F. Long and S. Chen and S. Wang and J. Jing},
  title = {Energy extraction from a Konoplya-Zhidenko rotating non-Kerr black hole},
  journal = {Nucl. Phys. B},
  year = {2018},
  volume = {926},
  pages = {1}
}

@article{KZ7,
  author = {F. Long and S. Wang and S. Chen and J. Jing},
  title = {Magnetic reconnection and energy extraction from a Konoplya-Zhidenko rotating non-Kerr black hole},
  journal = {Eur. Phys. J. C},
  year = {2025},
  volume = {85},
  pages = {26}
}

@article{KZ8,
  author = {K. He and C. Yang and X. Zeng},
  title = {Optical appearance of the Konoplya-Zhidenko rotating non-Kerr black hole surrounded by a thin accretion disk},
  journal = {arxiv},
  year = {2025},
  pages = {2501.06778}
}

@article{Gra1,
  author = {S. E. Gralla and A. Lupsasca},
  title = {Lensing by Kerr black holes},
  journal = {Phys. Rev. D},
  year = {2020},
  volume = {101},
  pages = {044031}
}

@article{Him1,
  author = {E. Himwich and M. D. Johnson and A. Lupsasca and A. Strominger},
  title = {Universal polarimetric signatures of the black hole photon ring},
  journal = {Phys. Rev. D},
  year = {2020},
  volume = {101},
  number = {2},
  pages = {084020}
}

@article{WPC,
  author = {M. Walker and R. Penrose},
  title = {On quadratic first integrals of the geodesic equations for type $S^{(22)}$ spacetimes},
  journal = {Commun. Math. Phys.},
  year = {2001},
  volume = {118},
  pages = {265}
}

@book{Chandrasekhar,
  author = {S. Chandrasekhar},
  title = {The Mathematical Theory of Black Holes},
  year = {1985},
  publisher = {Oxford University Press}
}
\bibliographystyle{unsrt}



\end{document}